\documentclass[preprint]{revtex4-1}
\usepackage{graphicx}
\usepackage{pstricks}

\begin{document}

\title{Tuning electronic and optical properties of bismuth monolayers by molecular adsorption}
\author{Erika N. Lima}
\affiliation{Universidade Federal de Rondon\'opolis, ICEN, 78725901 Rondon\'opolis, Mato Grosso,  Brazil}
\author{A. L. da Rosa}
\affiliation{Bremen Center for Computational Materials Science, University of Bremen, Am Fallturm 1, 28359 Bremen, Germany}
\affiliation{Universidade Federal de Goi\'as, Institute of Physics, Campus Samambaia, 74960600 Goi\^ania, Brazil}
\author{R. B. Pontes}
\affiliation{Universidade Federal de Goi\'as, Institute of Physics, Campus Samambaia, 74960600 Goi\^ania, Brazil}
\author{M. C. Silva}
\affiliation{Bremen Center for Computational Materials Science, University of Bremen, Am Fallturm 1, 28359 Bremen, Germany}
\affiliation{Max-Planck-Institut, Hamburg, Germany}
\author{T. Schmidt}
\affiliation{Instituto de F\'isica, Universidade Federal de Uberl\^andia, 38400-902, Uberl\^andia, MG, Brazil.}
\author{T. Frauenheim}
\affiliation{Bremen Center for Computational Materials Science, University of Bremen, Am Fallturm 1, 28359 Bremen, Germany}
\affiliation{Shenzhen Computational Science and Applied Research Institute, Shenzhen, China}
\affiliation{Beijing Computational Science Research Center, Beijing, China}

\begin{abstract}

  We perform first-principles calculations of electronic and dielectric properties of
  bismuthene functionalized with small ligands using first-principle
  calculations. We show that all functionalized structures have
  topological insulating (TI) behavior with a sizeable gap by calculating the Z$_2$ topological
  invariant. Furthermore the adsorption of all groups induce a quasi-planar structure to
  the initially pristine bismuthene structure. Finally we show that
  the dielectric properties show a large anisotropy with two main in plane absorption peaks.

\end{abstract}


\maketitle

\section{Introduction}

Several investigations on optical properties of two-dimensional materials primarily focuses
on graphene\,\cite{Geim:07,Novoselov:04}, hexagonal boron
nitride\,\cite{NatMat:2019}, transition metal dichalcogenides
(TMDCs)\,\cite{TMDC2018} More recently, other two-dimensional materials such
as bismuthene have also attracted the attention for promising
optoelectric applications. free-standing two-dimensional bismuthene
has been realized experimentally\,\cite{NatComm2020}. This kind of
structure demonstrate electrocatalytic efficiency for CO2 reduction
reaction.  Moreover, bismuth single-monolayer has ben sinthesized on
SiC substrates\,\cite{Reis2017}.  Previously, bismuth monolayers have
been predicted to be a 2D topological insulator
(TI)\,\cite{Kou:NL,Rivelino2015,AsiaNat,Reis2017}. These TI have 2D
edge states that are robust against applied strain and under
application of an external electric
field\,\cite{Kou:NL,MingYang2017}. Moreover, a sizeable band gap due
to Quantum Spin Hall (QSH) and valley-polarized quantum anomalous Hall
states have been predicted to emerge as the bismuth layers films are
adsorbed with hydrogen and other small functional
groups\,\cite{Kou:NL,Niu2015}.

A known problem has been the stability of these two-dimensional layers against
changes in the environment. Recently a route for creating
organic-terminated germanane via deintercalation of layered precursor
CaGe2 phases with organohalides has been
created\,\cite{JiangNT:2014,ProgressReports}. In
bismuthene there is still a lack of experimental investigations to
understand the stability of these systems in air and other atmosphere,
but functionalized bismuth layers with small radicals were also shown
to have topologically protected states in the band
gap\,\cite{Reis2017,Rivelino2015,Kou:NL}.

Angle-resolved photoemission spectroscopy (ARPES) measurements
observed Dirac-like band dispersions in $\alpha$-bismuthene. The
results enlighten the search of nonsymmorphic 2D materials for optical
applications\,\cite{ACS2020}. Density-functional theory calculations
of optical properties of single-layer buckled bismuthene found that a
direct band gap can be obtained by applying a 5\% strain. However, the
optical properties of bismuthene do not depend significantly on the
applied strain\,\cite{Bi:2020}.  Therefore, understanding of the
optical properties of functionalized bismuthene layers can render comprehensive
understanding of bismuthene possible applications under strain or
reactive environment. In this paper we perform first-principles
calculations of electronic and dielectric properties of bismuthene
functionalized with small ligands containing hydrogen and carbon. We
show that the functionalized structures exhibit two-dimensional
topological insulating behavior with a sizeable bandgap. This is
confirmed by calculating the Z$_2$ topological invariant of the hybrid
systems. As a general feature the adsorption of all groups induce a
planar structure to the initially pristine buckled bisuthene. Finally
we show that the dielectric properties show a large anisotropy, making
this material suitable for novel optoelectronic devices.

\section{Methodology}

In this work we use density functional
theory\,\cite{Hohenberg:64,Kohn:65} within the generalized gradient
approximation\,\cite{Perdew:96} as implemented in the VASP
package\,\cite{Kresse:99,Shishkin:07} to investigate the electronic
structure of functionalized bismuthene. The projected augmented wave
method (PAW)\,\cite{Bloechel:94,Kresse:99} has been used. A
($1\times$1) supercell with a (8$\times$8$\times$1) {\bf k}-point
sampling and an energy cutoff of 500\,eV is used to calculate the
atomic relaxation and electronic structure of the functionalized
layers. The calculations of the dielectric properties using GW
method\,\cite{Shishkin:07,Hedin:65} were performed using a
$(6\times6\times1)$ {\bf k}-point mesh and an energy cutoff of
300\,eV. All calculations include spin-orbit coupling. Atomic forces
were converged until 10$^{-4}$\,eV/{\AA}. Z$_2$ calculations with the Z2Pack
package\,\cite{Z2Pack1,Z2Pack2}.

\section{Results and Discussions}

The stability of the buckled and planar structure has reported in our
previous publication\,\cite{JPCC2020} The free-stading buckled bismuthene
was found to be more stable than the flat one with in-plane lattice
constant of 5.30\,{\AA}.  Some routes for tuning the electronic
structure of two-dimensional materials have been proposed. A promising
one is the surface modification of 2D layers with small organic
molecules or functional groups \,\cite{ProgressReports,JiangNT:2014,Kou:NL}.
We have therefore functionalized the bare bismuth
layers by adsorption of small ligands, namely -H, -CH$_3$, -C$_2$H,
-CH$_2$CHCH$_2$ and -CH$_2$OCH$_3$. We have considered a coverage of
1\,ML (monolayer) with the ligands sitting on the top positions of
bismuth atoms on both sides of the bilayer, mimicking a full covered
material, with grafting on almost each Bi atoms. The choice for this
particular coverage is based on experimental results for grafting
germanium layers\,\cite{JiangNT:2014}. 

\begin{figure}[ht!]
  \pspicture(16,8)
  \centering
  \begin{tabular}{cc}
\rput(2,6){\includegraphics[width=4cm,scale=1.0,clip, keepaspectratio]{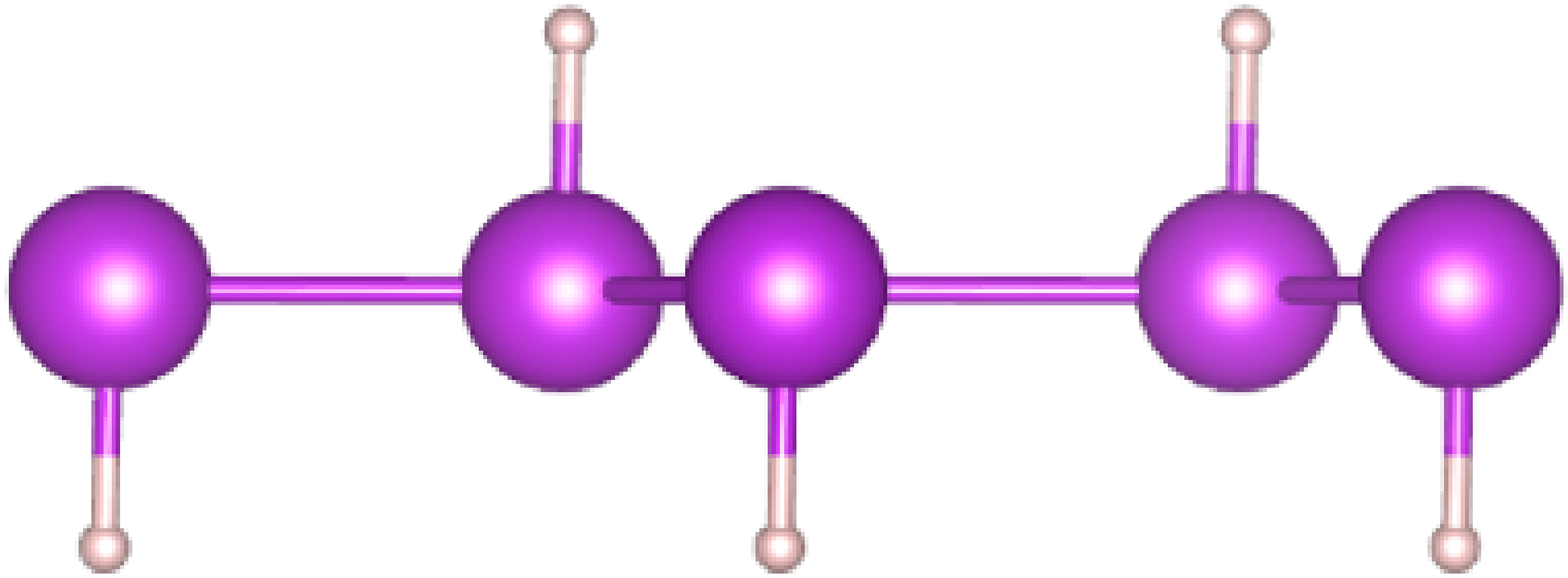}}
\rput(7,6){\includegraphics[width=4cm,scale=1.0,clip,keepaspectratio]{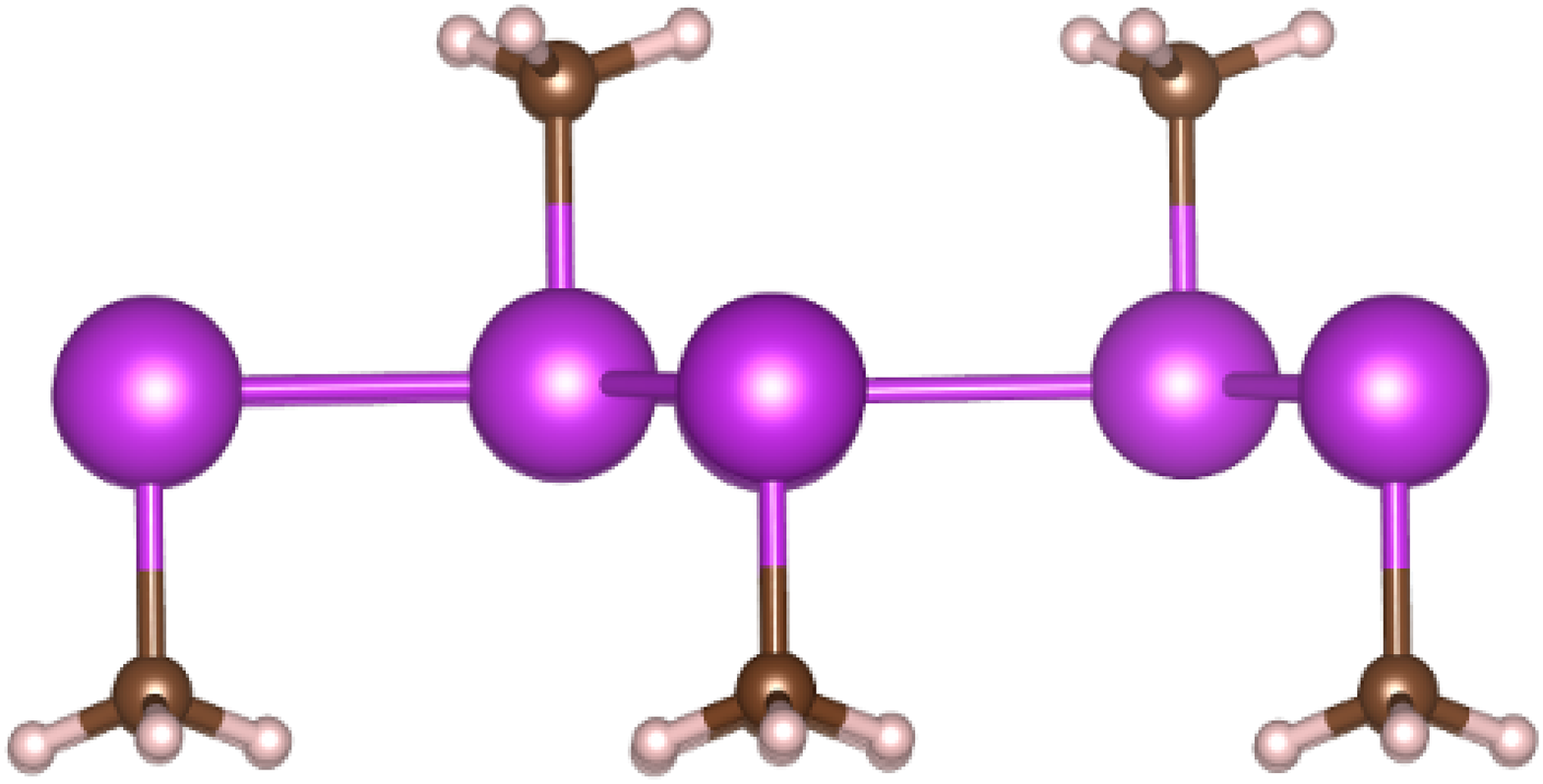}}
\rput(12,6){\includegraphics[width=4cm,scale=1.0,clip,keepaspectratio]{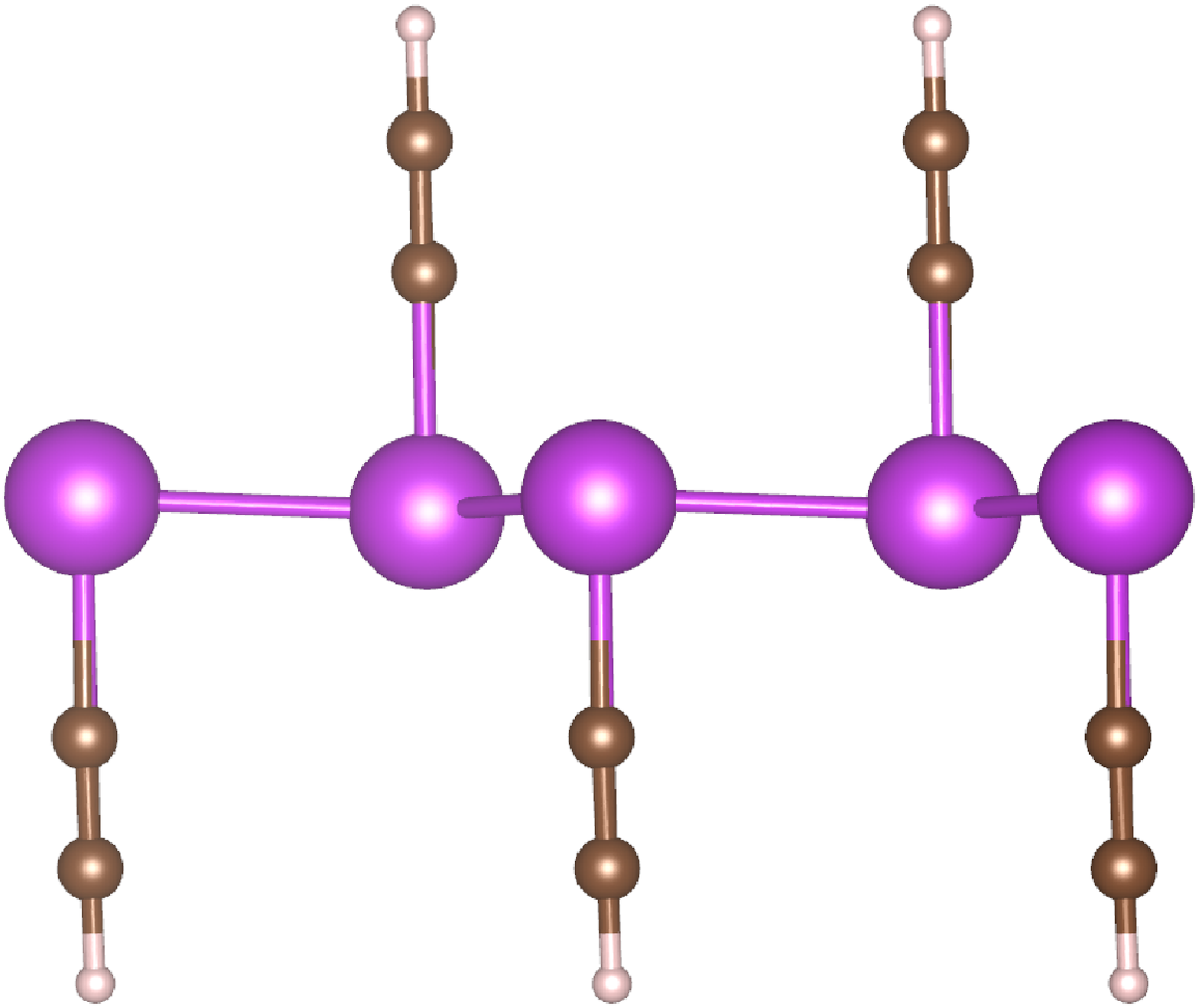}}
\rput(2,2){\includegraphics[width=4cm,scale=1.0,clip,keepaspectratio]{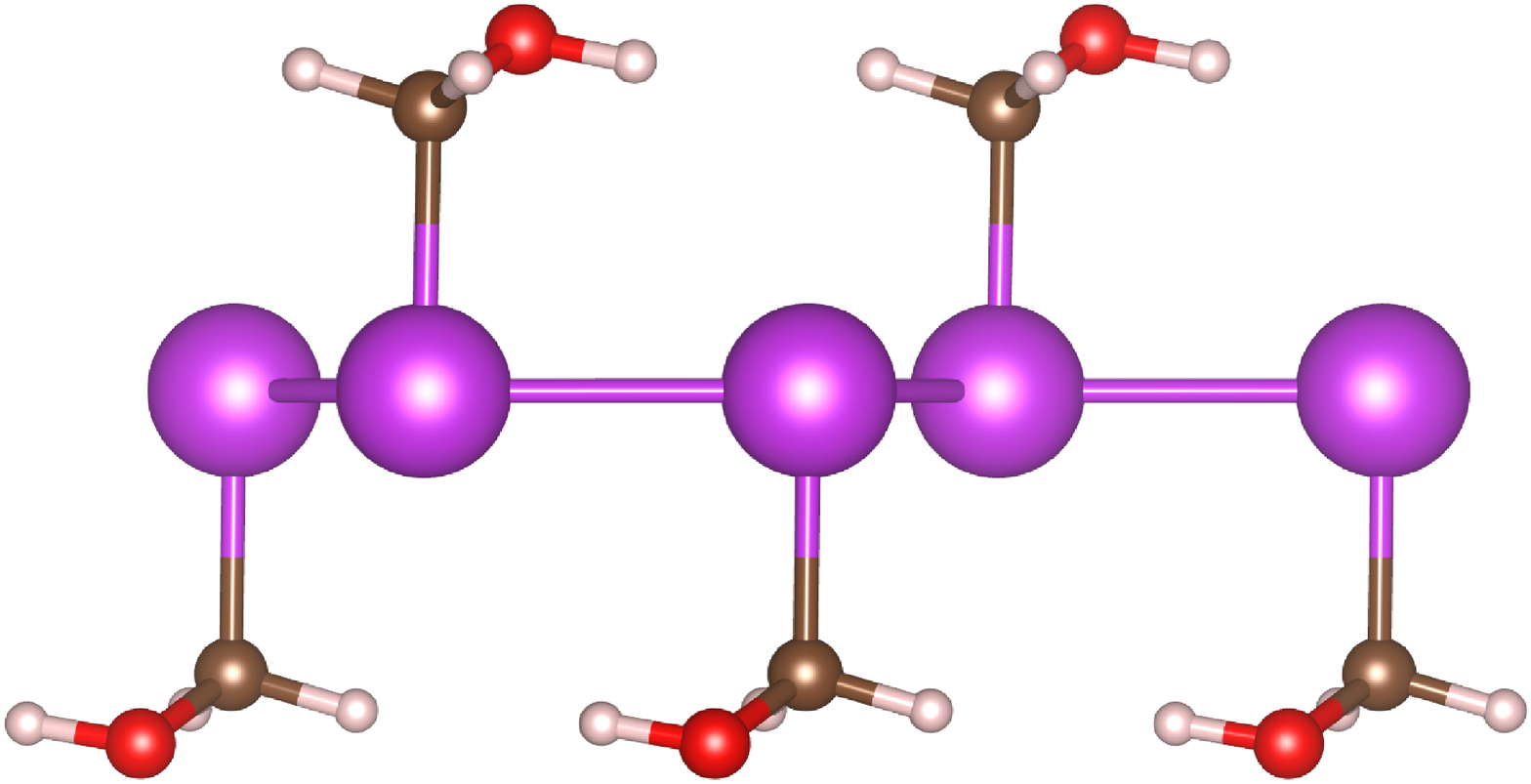}}  
\rput(7,2){\includegraphics[width=4cm,scale=1.0,clip, keepaspectratio]{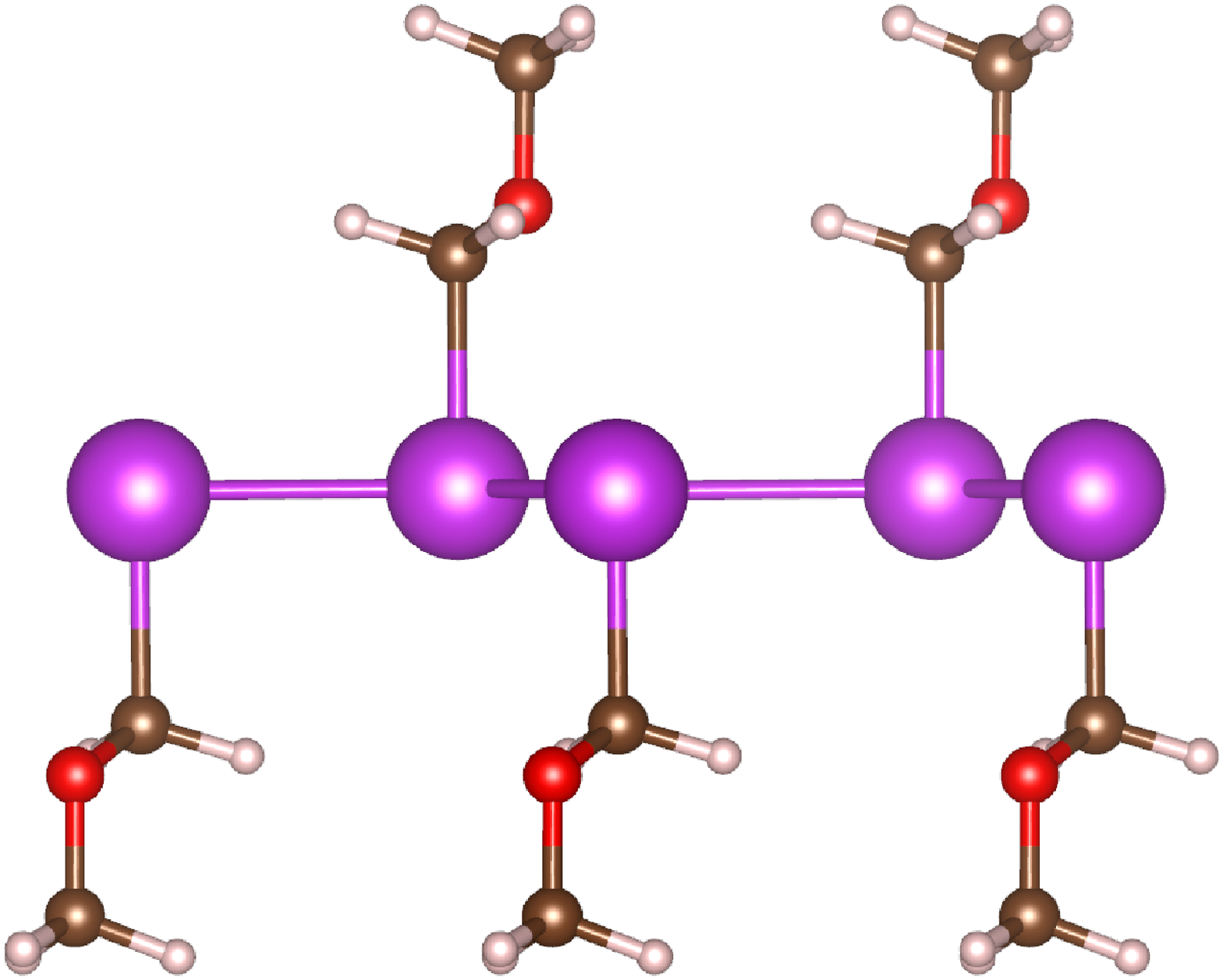}}
\rput(12,2){\includegraphics[width=4cm,scale=1.0,clip,keepaspectratio]{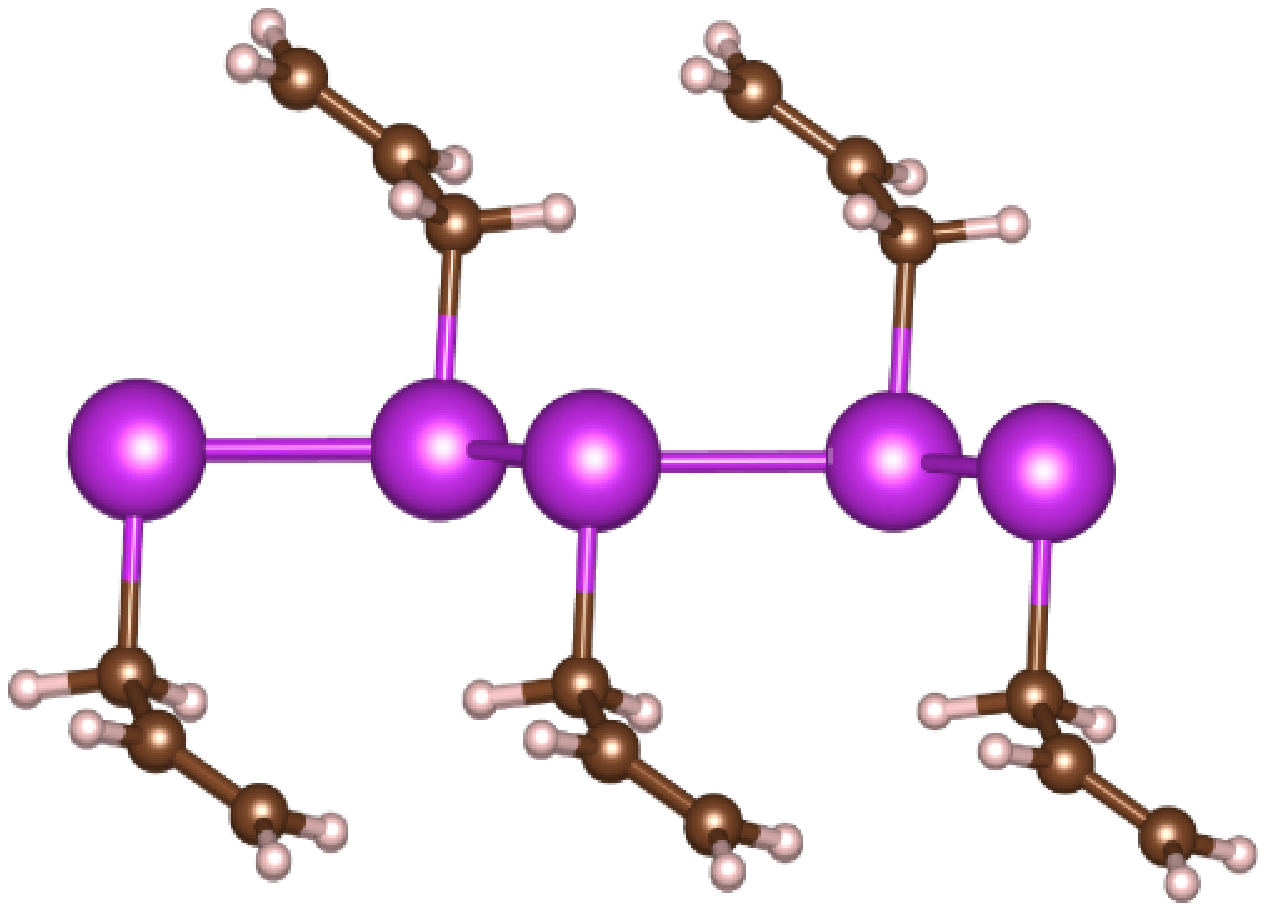}}
\rput(0,8){a)}
\rput(5,8){b)}
\rput(10,8){c)}
\rput(0,3){d)}
\rput(5,3){e)}
\rput(10,3){f)}
\end{tabular}
\endpspicture
\caption{\label{fig:structures}(color online) Side views of functionalized bismuth layers with ligands. a) -H, b) -CH$_3$, c) -C$_2$H, d) -CH$_2$OH, e)  -CH$_2$OCH$_3$ and f) -CH$_2$CHCH$_2$.} 
\end{figure}

Configurations where the radicals are on top positions are more stable
than hollow and bridge sites, since starting from bridge/hollow sites sponteanously relax to top positions.

Fig.\,\ref{fig:structures}(a) shows the bismuth
layers functionalized with hydrogen. The H-Bi bond lenght is
1.83\,{\AA}.  The lattice parameter is 5.54\,{\AA}, slightly larger
than the lattice parameter of the metastable planar pure bismuthene.
As it can be seen, the structure has a relaxed planar configuration.
Fig.\,\ref{fig:structures}(b) shows the bismuth layers under
adsorption of -CH$_3$. The radical binds to the bismuth surface
through the molecule carbon atoms. The C-Bi distance is 2.30\,{\AA}
and the lattice parameter is 5.49\,{\AA}, also similar to the planar
pristine structure. We notice that this structure is nearly planar,
with a very small buckling. Fig.\,\ref{fig:structures}(c) shows the
bismuth layers functionalized with -CH$_2$. This structure relax to a
near planar structure.  This configuration leads to an in-plane
lattice parameter of xx\,{\AA}, which further confirmed that the main
driving force for the planarity is due to the ligand-ligand
interaction.

Fig.\,\ref{fig:structures}(e) shows the relaxed structure for the
${\rm -CH_2CHCH_2}$. The ligand-ligand interaction is important since
due to steric effects, the ligand try to minimize their repulsion
inside the unit cell by assuming a tilted mode. The C-Bi distance is
2.17\,{\AA} and the in-plane lattice parameter is 5.50\,{\AA}. Finally
we show the adsorption of ${\rm -CH_2OCH_3}$ in
Fig.\,\ref{fig:structures}(f). The relaxation is very similar to the
previous discussed structure, with the ligand adsorbed in a tilted
configuration. The C-H bond length is 2.39\,{\AA} and the lattice
parameters is 5.55\,{\AA}.  We should notice that water does not
dissociate on Bi(111) surfaces\,\cite{Lust}. Therefore the -OH groups
are not stable on bismuth layers. Fig.\,\ref{fig:structures}(a) shows
the bismuth layers functionalized with hydrogen.

\begin{figure}[ht!]
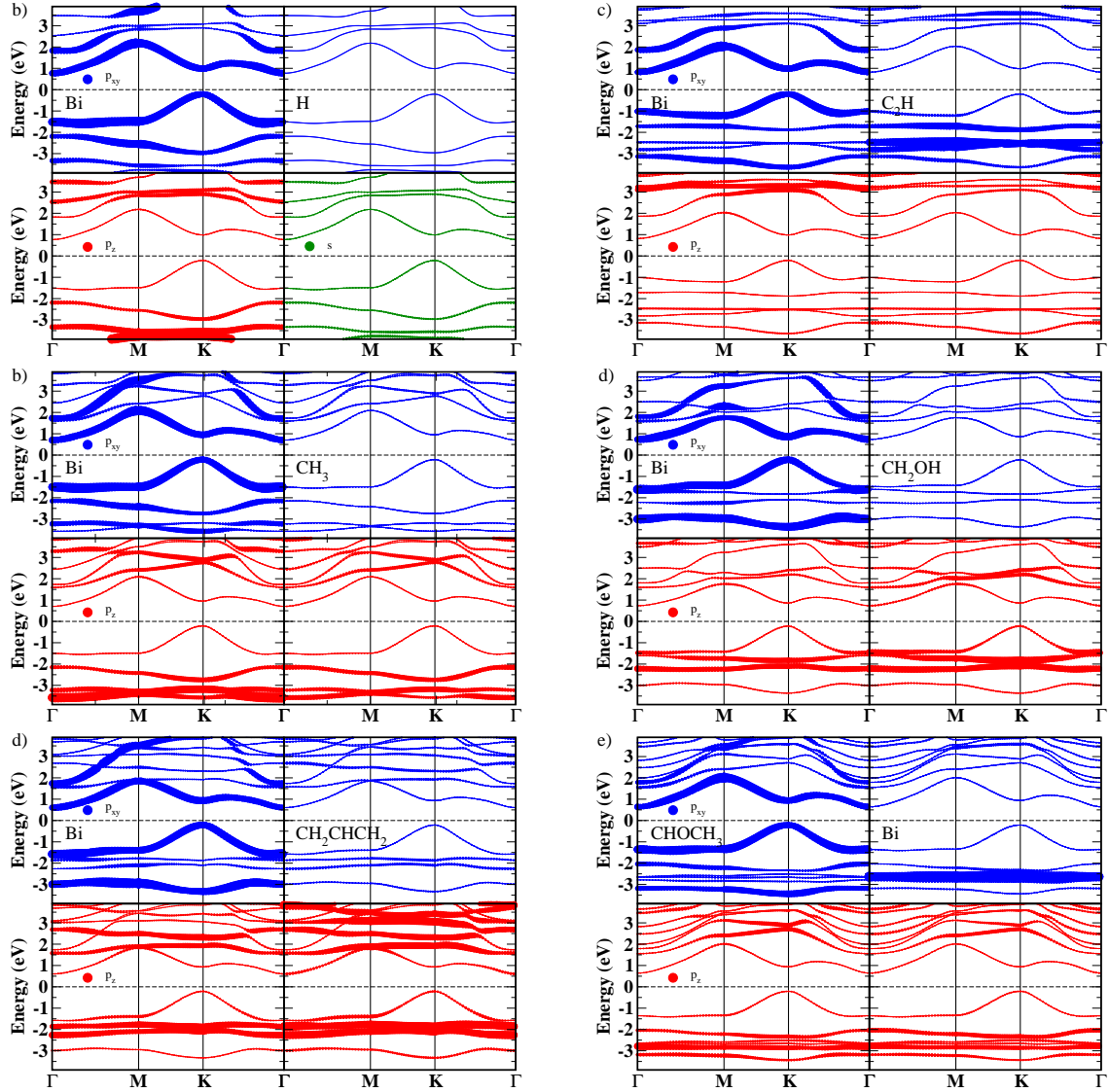

\pspicture(16,16)
\rput(4,12){\includegraphics[width = 7cm,scale=1, clip = true]{./bismuthene_H_projected_bands.eps}}
\rput(12,12){\includegraphics[width = 7cm, scale=1, clip = true]{./bismuthene_C2H_projected_bands.eps}}
\rput(4,7){\includegraphics[width = 7cm,scale=1, clip = true]{./bismuthene_CH3_projected_bands.eps}}
\rput(12,7){\includegraphics[width = 7cm, scale=1, clip = true]{./bismuthene_CH2OH_projected_bands.eps}} 
\rput(4,2){\includegraphics[width = 7cm,scale=1, clip = true]{./bismuthene_CH2CHCH2_projected_bands.eps}}
\rput(12,2){\includegraphics[width = 7cm,scale=1, clip = true]{./bismuthene_CH2OCH3_projected_bands.eps}}
\endpspicture
\caption{Projected band structure of surface modified bismuthene structures. Only
  significant orbital contributions are shown for bismuthene adosrbed with a) -H, b) -CH$_3$, c) -C$_2$H, d) -CH$_2$OH, e) -CH$_2$OCH$_3$ and f) -CH$_2$CHCH$_2$.}
\label{fig:projected_bands}
\end{figure}

In order to further understand the individual orbital contributions to
the band edges, we have calculated the orbital projected band
structures as shown in Fig. \,\ref{fig:projected_bands}.

The electronic structure of bare bismuthene layers is shown in
Fig.\,\ref{fig:band_bare} The top of the valence band (VBM) is at the
M point. The band gap is 0.79\,eV (direct) for the planar geometry and
0.49\,eV (indirect) for the buckled geometry. The band crossing in the
buckled structure occurs at 0.15\,eV below the VBM, giving origin to
the Rashba effect.  The spin-orbit coupling splits the double
degenerate bands for each spin channel at the VBM. The decomposed band
structure shows that the bottom of the conduction band has mainly
$p_x$ and $p_y$ character. The VBM has mixed $p_y$ and $p_z$
characters along the $\Gamma$-K and $p_z$ and $p_x$ characters along
$\Gamma$-M. Our results agree very well with previous
ones\,\cite{MingYang2017,Kou:NL}.

Now we discuss the ligand-adsorbed layers. An overall picture is that
the $p_z$ component completely vanishes at the $\Gamma$ point leading
to a band structure very similar to the planar bismuthene, but now
with indirect gaps. This suggests that the strain induced by these
ligands can not only stabilize the bismuth layers but also keeps the
semiconductor character, although the band gap turns to be
indirect.

\begin{table}
\begin{center}
\caption{\label{table:gaps} Topological invariant Z$_2$, electronic band gaps, main absoprtion peak positions and lattice parameter $a$ of bare and functionalized bismuthene within GGA-PBE. (i) means an indirect and (d) a direct band gap. In brackets the relative positions to the buckled bare bismuthene absorption spectrum are shown.}
\begin{tabular*}{16cm}{@{\extracolsep{\fill}}lcccccc}
\hline
structure                  & Z$_2$  & classification & $a(\AA)$   & band gap (eV) & 1${\rm st}$ peak & 2${\rm nd}$ peak \\
bismuthene (buckled)       & 1             & TI            & 4.34 & 0.49 (i) &   4.2 & 7.3 \\
bismuthene (planar)        & 0             & trivial       & 5.30 & 0.79 (d) &       &         \\
bismuthene-H               & 1             & TI            & 5.54 & 0.99 (i) &   2.3 (1.9) & 3.7 (3.6) \\
bismuthene-CH$_3$          & 1             & TI            & 5.49 & 0.93 (i) &   2.0 (2.2) & 3.5 (3.8) \\
bismuthene-C$_2$H          & 1             & TI            & 5.53 & 1.03 (i) &   2.4 (1.8) & 3.8 (3.5) \\
bismuthene-CH$_2$OH        & 1             & TI            & 5.53 & 0.94 (i) &       &    \\
bismuthene-CH$_2$OCH$_3$   & 1             & TI            & 5.55 & 0.66 (i) &   2.1 (2.1) & 3.2 (4.1)  \\
bismuthene-CH$_2$CHCH$_2$  & 1             & TI            & 5.70 & 0.49 (i) &       &    \\ 
\hline
\end{tabular*}
\end{center}
\end{table}

In order to investigate whether the bismuth hybrid layers behave like
topological insulators we have calculated the Z$_2$ structures
according to Ref.\,\cite{Z2}. The results are shown in Table
\ref{table:gaps}.  All functionalized structures are topological
insulators with indirect band gap. Besides the bare buckled structure
is also TI. The planar structure has trivial, direct gap. As the
lattice parameter increases, the band gap decreases. 

In order to understand further the individual otbital contributions to
the band edges, we have calculated the projected band structures on
the orbitals shown in Figs. \,\ref{fig:projected_bands} (a)-(f).  For the Bi-H (bismuthane) the VBM has
both $p_x$ and $p_y$ characters belonging to bismuth layers as shown
in Fig.\,\ref{fig:projected_bands} (a). The hydrogen states lie deeper
in the valence band (between -4 and -2\,eV) and therefore there is a
negligible overlap with Bi orbitals. The CBM has also $p_x$ and $p_y$
character which belong to bismuth layers. The band gap is 0.99\,eV and
indirect as shown in Table\,\ref{table:gaps}. The VBM of the
methyl-functionalized structure, Bi-CH$_3$, has mostly $p_x$ and $p_y$
characters also belonging to bismuth layers underneath. The
hybridization of the carbon states belonging to the radical at the VBM
is small, so that the main character comes from the bismuth
layers. The CBM retains the $p_x$ and $p_y$ character related to the
pure bismuth layers, shown in Fig.\ref{fig:projected_bands} (b). The
band gap is 0.93 and indirect.

The ${\rm Bi-CH_2CHCH_2}$ band structure shown in
Fig.\ref{fig:projected_bands} (c). Again the overlap of Bi orbitals
with C orbitals is negligible.  The band gap is indirect and smaller
than the previous discussed structures and has a value of
0.83\,eV. The Bi-CH$_2$OCH$_3$ band structure  shown in
Figs.\,\ref{fig:projected_bands} (d) has a hybridization of the ligand C-$p$ and O-$p$ orbitals and Bi-$p$ orbitals around 1.0\,eV below the valence band,
although the VBM has a large contribution from the bismuth layers. The
band gap is indirect and smaller than the previous discussed
structures and has a value of 0.66\,eV. This structure shows a similar
behavior to the ${\rm Bi-CH_2CHCH_2}$ and indirect transition from
VBM to CBM.

An overall picture is that the $p_z$ component completely disappears
at the $\Gamma$ point for the functionalization of bismuth layers with
-H, ${\rm -CH_3}$ and ${\rm -CH_2CHCH_2}$ or is drastically reduced in
the case of the ${\rm -CH_2OCH_3}$ adsorption, leading to a band
structure very similar to the planar bismuthene, but with indirect
gaps (the planar pristine has a direct gap at the M point). it
suggests that the strain induced by these ligands can not only
stabilize the bismuth layers keeping their topological behavior but
also almost recover the direct band gap. Therefore, we conclude that
the band gap can be tunes by changing the size and electronic affinity
of the ligands.

Single-layer two-dimensional (2D) materials display strong
electron–photon interactions that could be utilized in efficient light
modulators on extreme subwavelength scales.  The optical response
giving by the calculation of the dielectric function is determined
then from infra-red to deep-ultraviolet energies.  The imaginary part
of the dielectric function is calculated directly from the electronic
structure through the joint density of states and the matrix elements
of the momentum, \textbf{p}, between occupied and unoccupied
eigenstates according to:

\begin{equation}
\epsilon_{2}^{ij}(\omega)= {4\pi^2 e^2 \over \Omega m^2 \omega^2}
\sum_{{\bf k} n n^{\prime}}
\bigl\langle {\bf k} n \big | p_{i} \big | {\bf k} n^{\prime} \bigr\rangle
\bigl\langle {\bf k} n^{\prime} \big | p_{j} \big | {\bf k}
n \bigr\rangle \times f_{{\bf k}n}\, \bigl(1 - f_{{\bf k} n^{\prime}}\bigr) \,
\delta\bigl( E_{{\bf k} n^{\prime}} - E_{{\bf k} n} - \hbar \omega
\bigr).
\end{equation}

In this equation, {\it e} is the electron charge, {\it m} the electron mass,
$\Omega$ is the volume of the crystal, $f_{\bf k n}$ is the Fermi
distribution function and $\big |\bf k {\it n} \bigr\rangle$ is the crystal wave function corresponding to the ${\it n^{th}}$ eigenvalue
${\it E}_{{\bf k}n}$ with crystal wave vector {\bf k}.

\begin{figure}[ht!]
\begin{tabular}{cc}
\includegraphics[width = 8cm,scale=1, clip = true, keepaspectratio]{./diel_bi_pure_GW.eps}&
\includegraphics[width = 8cm,scale=1, clip = true, keepaspectratio]{./diel_h_bi_GW.eps}\\
\includegraphics[width = 8cm,scale=1, clip = true, keepaspectratio]{./diel_c2h_bi_GW.eps}&
\includegraphics[width = 8cm,scale=1, clip = true, keepaspectratio]{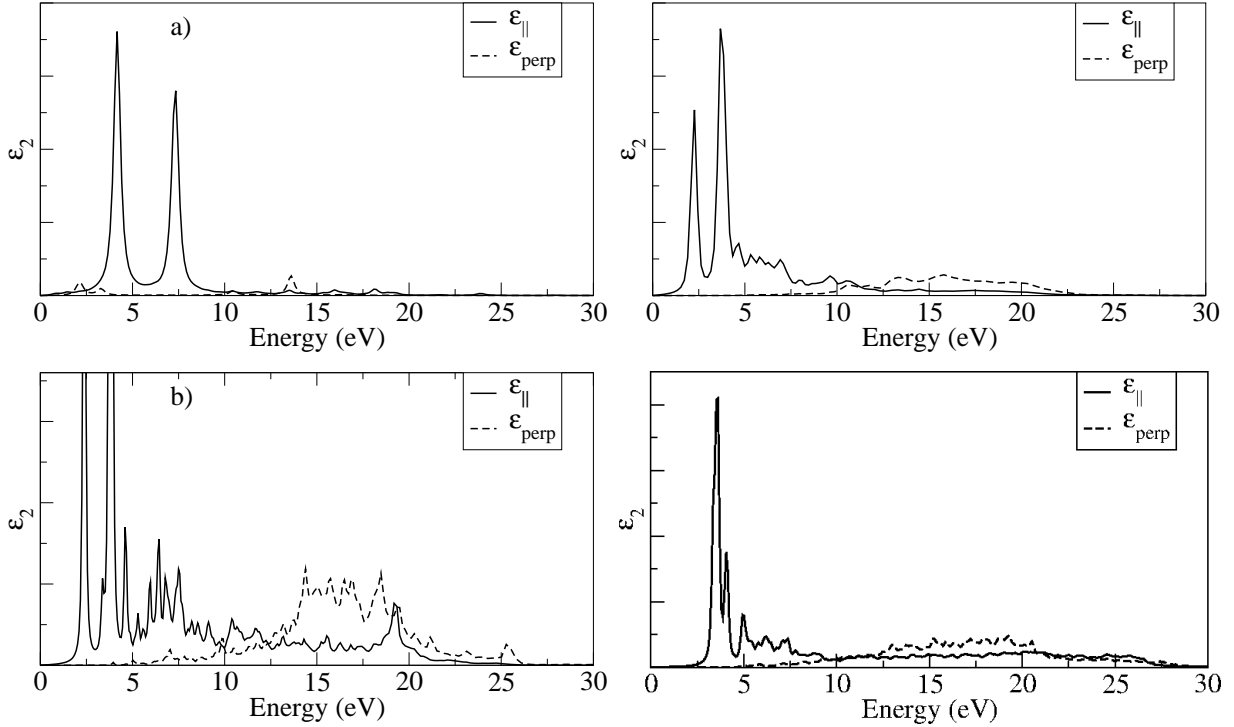}
\end{tabular}
\caption{Imaginary part of the dielectric function $\epsilon_2(\omega)$ for bare
  and ligand modified bismuthene calculated within the GW approximation. a) bismuthene, b) bismuthane, c) C$_2$H and d) ${\rm CH_2OCH_3}$.  The average of the parallel $\epsilon_{||}=(\epsilon_{\rm xx}+\epsilon_{\rm yy})/2$ and
perpendicular ${\rm \epsilon_{perp}=\epsilon_{zz}}$ components are shown.}
\label{fig:diel}
\end{figure}

In anisotropic materials, dielectric properties must be described by
the dielectric tensor.  In Fig. \ref{fig:diel} we show the average of
the parallel $\epsilon_{|}=(\epsilon_{xx}+\epsilon_{yy)})/2$ and
perpendicular $\epsilon_{perp}=\epsilon_{zz}$ directions of light
polarization for $\varepsilon_{2}(\omega)$ using GW. Besides the
results for bare bismuthene we show three showcases, two for a
small ligands -C$_2$H and CH$_3$ for a larger ligand ${\rm -CH_3OCH_2}$. 

 We identify a large anisotropy for bare and functionalized bismuth
 monolayers. This indicates a strong bismuth-bismuth hybridization
 within the basal plane, but smaller overlap perperdicular to the
 plane.  Fig.\,\ref{fig:diel}(a) shows the results for bare
 bismuthene.  The spectrum can be roughly divided into two main
 regions, around 4.2 and 7.3\,eV. The $\epsilon_{\rm perp}$ is
 basically supressed. Upon adsorption of -H shifts of 2.3 and 3.7 eV
 as seen in Fig.\,\ref{fig:diel}(d) . Upon adsorption of ${\rm -C_2H}$
 groups, as shown in Fig.\,\ref{fig:diel}(c), peaks around 2.4 and 3.8
 eV are seen. Another broad region between 6.0 and 8.0 eV is also
 presnet. There is a red shift implying that the system forms
 bonds. The ${\rm -CH_2OCH_3}$ shown in Fig.\,\ref{fig:diel}(d) has
 defined peaks centered around 2.1 and 3.2\,eV. The general trend is
 the appearence of two distinct peaks. In functionalized bismuthene
 these two peaks appear are red-shift with respect to bare buckled
 bismuthene. The As the lattice constant increases, the spectrum is
 shifted to left, meaning that there is an interplay between size and
 reacitivity, as suggested for small ligands in
 germanene\,\cite{JiangNT:2014}. Similarly, anisotropic and
 thickness-dependent optical properties of a two-dimenasiona layered
 monochalcogenide of germanium sulfide\,\cite{GeS} and black
 phosphorous\,\cite{blackP} has also been reported.

\section{Conclusions}

All the functional groups interact weakly with the bismuth
surface. The transition from buckled to planar structure upon ligand
adsorption is mainly due to the molecule-molecule interaction rather
than the molecule substrate. This leads to a change in the electronic
structure, yielding all the investigated groups as topological
insulators. It is worth noting that lower coverages, where the
ligand-ligand intercation is weaker, do not lead to buckled
structures.  The stability of these planar structures is mainly due by
strain due to the ligand-ligand interaction rather than
ligand-substrate interaction, meaning that the band gap of such
systems can be tuned by chosing the appropriate ligand or molecule.

\section{Acknowledgements}

We acknowledge the financial support from the Brazilian Agency CNPq
and German Science Foundation (DFG) under the program FOR1616. The
calculations have been performed using the computational facilities of
Supercomputer Santos Dumont and at QM3 cluster at the Bremen Center
for Computational Materials Science and CENAPAD.

\bibliographystyle{apsrev}

\end{document}